\documentclass[12pt]{article}
\usepackage{graphicx}

\newcommand\pubdate{March 14, 2014}

\def\napoli{Instituto de F\'{\i}sica, Universidade de S\~ao Paulo,
C.\ P.\ 66.318, 05315-970 \\ S\~ao Paulo, Brasil} 
\def\support{\footnote{Work supported in part by Grant-in-Aid for Scientific Research No. 23540315, Japan Society for the Promotion of Science.}}

\def\Title#1{\begin{center} {\Large #1 } \end{center}}
\def\Author#1{\begin{center}{ \sc #1} \end{center}}
\def\Address#1{\begin{center}{ \it #1} \end{center}}

\newcommand\pubblock{\rightline{\begin{tabular}{l} 
         \pubdate  \end{tabular}}}
\newenvironment{Abstract}{\begin{quotation}  }{\end{quotation}}
\newenvironment{Presented}{\begin{quotation} \begin{center} 
             EXPANDED FROM A TALK AT\end{center}\bigskip 
      \begin{center}\begin{large}}{\end{large}\end{center} \end{quotation}}
\def\Acknowledgements{\bigskip  \bigskip \begin{center} \begin{large}
             \bf ACKNOWLEDGEMENTS \end{large}\end{center}}

\def\beq{\begin{equation}}
\def\eeq#1{\label{#1}\end{equation}}
\def\eeqn{\end{equation}}

\def\beqa{\begin{eqnarray}}
\def\eeqa#1{\label{#1}\end{eqnarray}}
\def\eeqan{\end{eqnarray}}

\let\bar=\overbar

\def\Dslash{\not{\hbox{\kern-4pt $D$}}}
\def\dslash{\not{\hbox{\kern-2pt $\del$}}}

\def\msb{{\bar{\ssstyle M \kern -1pt S}}}

\begin{document}
\begin{titlepage}
\pubblock

\vfill
\Title{Neutrino Physics Now and in the Near Future}
\vfill
\Author{ Hisakazu Minakata\support}
\Address{\napoli}
\vfill
\begin{Abstract}
The current status of neutrino physics is reviewed with some near future perspective. After recollecting the birth of modern neutrino physics with nonzero masses and flavor mixing, I summarize the present status of measurement of the mixing parameters in 2-3, 1-2, and 1-3 sectors of the MNS matrix. Then, I describe the attempts to uncover the regularities, if any, in the measured values of the mixing angles; mostly reviewing. Yet, a possible large deviation of $\theta_{23}$ to the second octant may trigger interests in the triangle relation of the lepton mixing angles. In the latter part of my lecture some perspective of determination of the mass hierarchy and measurement of lepton Kobayashi-Maskawa phase $\delta$ are described. Finally, I discuss the prospects of the new, fast developing field of high-energy neutrino astrophysics, and the emerging new precision era of cosmology and particle physics. I conclude with optimistic speculations.

\end{Abstract}
\vfill
\begin{Presented}
10th International Symposium on Cosmology and Particle Astrophysics (CosPA 2013), Honolulu, Hawaii, USA, November 12-15, 2013
\end{Presented}
\vfill
\end{titlepage}
\def\thefootnote{\fnsymbol{footnote}}
\setcounter{footnote}{0}

\section{Introduction}

There is a great disparity in the atmosphere in our community between {\em now} and the time Kamiokande II experiment claimed ``atmospheric neutrino anomaly''  \cite{Kam-II} to which very few number of people coined.\footnote{
The current stage of dark matter search might have some similarities with the era in the sense that no positive experimental claim does not appear to become the consensus in the community. The clear differences is that people are confident that the dark matter exists. It would be extremely interesting to see what outcome emerges out of it.
}
Nowadays, people say {\em ``of course, neutrinos have masses and they mix''}, but the recognition has been pioneered by the extensive efforts to confirm the anomaly, which finally revealed the phenomenon of neutrino oscillation \cite{SKatm-evidence}. It is this result that made possible the above {\em emphasized} statement. 
Fortunately, it did not take so long time after the discovery that long wavelength solar scale oscillation is uncovered by two entirely different type of experiments, KamLAND reactor neutrino experiment \cite{KamLAND} and the solar neutrino observation \cite{solar-review}. It may be fair to say that for the latter the final stone has been placed by SNO with the help by SK \cite{SNO-SK}, and by SNO itself in {\em in situ} manner \cite{SNO02,SNO-3phase}.
Thus, the three-flavor mixing scheme of neutrinos is established; A brave theoretical suggestion for lepton flavor mixing \cite{MNS} became the reality. 
For more detailed account of story of the solar, the atmospheric and the reactor experiments, see e.g., \cite{solar-review,atm-review,reactor-review}.\footnote{
I must first apologize that my citation of the references in this report is largely arbitrary, but I also remind you that the proper one costs $\sim 100$ pages.
}

\section{All the mixing angles are measured}

At the CosPA 2013 conference, I was very happy to convey the message to friends in astroparticle and cosmology communities that we finally completed our understanding of the three-flavor neutrino mixing in the sense that all the mixing angles required in the scheme are measured. Let me briefly describe the present status for you, with some arbitrary comments. Note that the data are updated after CosPA.

Let us start by defining the lepton flavor mixing matrix, the MNS matrix $U_{MNS}$ \cite{MNS}, which relates the neutrino flavor eigenstate $\nu_{\alpha}$ ($\alpha = e, \mu, \tau$) to the mass eigenstate $\nu_{i}$ ($i=1, 2, 3$) as $\nu_{\alpha} = \left( U_{MNS} \right)_{\alpha i} \nu_{i}$. In its standard parametrization it takes the form $U_{MNS} = U_{23} U_{13} U_{12} U_{\rm phase}$ where 
\begin{eqnarray}
U_{MNS}^{\rm Dirac} = U_{23} U_{13} U_{12} &=&
\left[
\begin{array}{ccc}
1 & 0 &  0  \\
0 & c_{23} & s_{23} \\
0 & - s_{23} & c_{23} \\
\end{array}
\right] 
\left[
\begin{array}{ccc}
c_{13}  & 0 &  s_{13} e^{- i \delta}   \\
0 & 1 & 0 \\
- s_{13} e^{ i \delta}  & 0 & c_{13}  \\
\end{array}
\right] 
\left[
\begin{array}{ccc}
c_{12} & s_{12}  &  0  \\
- s_{12} & c_{12} & 0 \\
0 & 0 & 1 \\
\end{array}
\right], 
\nonumber \\
U_{\rm phase} &=&
\left[
\begin{array}{ccc}
1 & 0 &  0  \\
0 & e^{ i \alpha_{21} } & 0\\
0 & 0 & e^{ i \alpha_{31} } \\
\end{array}
\right], 
\label{MNSmatrix}
\end{eqnarray}
where $c_{12} \equiv \cos \theta_{12}$, $s_{12} \equiv \sin \theta_{12}$ etc., $\delta$ stands for the lepton Kobayashi-Maskawa (KM) phase \cite{KM}. $U_{\rm phase}$ is the phase matrix with the Majorana phases $\alpha_{21}$ and $\alpha_{31}$ which exists if the neutrinos are Majorana particle.

\subsection{2-3 sector parameters $\Delta m^2_{32}$ and $\theta_{23}$}
\label{sec:2-3}

\begin{figure}[htb]
\centering
\vglue 2mm
\includegraphics[width=0.98\textwidth]{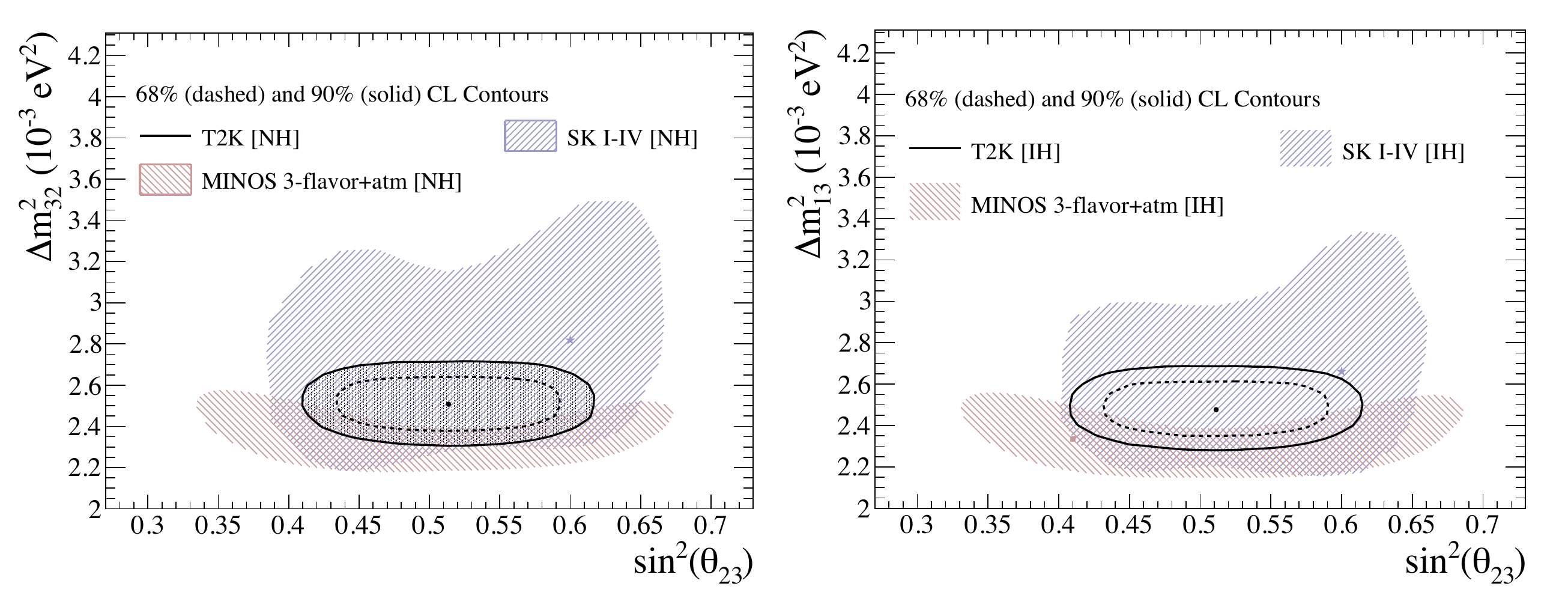}
\caption{
The regions in $\sin^2 \theta_{23} - \Delta m^2_{32}$ space allowed by the long-baseline accelerator neutrino experiments T2K and MINOS, as well as the one by the Super-K atmospheric neutrino data. 
This figure is taken from \cite{Minamino-KEK}.
}
\label{fig:23-parameters}
\end{figure}

The status of measurement of the 2-3 sector parameters changed in the last week by the new papers from T2K \cite{Abe:2014ugx} and MINOS \cite{Adamson:2014vgd}. With (more than) doubled statistics than that in \cite{Abe:2013fuq}, T2K says that the error of $\Delta m^2_{32}$ is $\pm$4.0\% for both the normal and the inverted mass hierarchies. Whereas MINOS reported that the error is 3.8\% for the normal and 4.3\% for the inverted mass hierarchies. These results together with the previous results of the Super-K atmospheric data (updated version of \cite{Wendell:2010md}) are shown in Fig.~\ref{fig:23-parameters}, which is taken from  \cite{Minamino-KEK}. 
As it stands, the accelerator experiments MINOS and T2K have advantage in accuracy of measuring $\Delta m^2_{32}$ because of the better $L/E$ resolution. 

To my knowledge this is the first time to see the accuracy of determination of $\sin^2 \theta_{23}$ by accelerator experiment surpasses (though not with wide margin) that of Super-K atmospheric data, which has been leading the race for more than 15 years. 
Yet, one may say that the error of $\sin^2 \theta_{23}$ is still large, $\simeq \pm 11\%$ ($1 \sigma$) \cite{Abe:2014ugx}, in comparison with the other mixing angles which will be summarized in the following two subsections.\footnote{
It is known that even though accuracy of measurement of $\sin^2 2 \theta_{23}$ is reasonably good, there are mainly two obstacles which prevent translation of the good accuracy to that of $\sin^2 \theta_{23}$ \cite{Minakata:2004pg}: 
(1) Jacobian from $\sin^2 2 \theta_{23}$ to $\sin^2 \theta_{23}$ is large at near maximal $\theta_{23}$, and (2) the two octant clones merge which leads to a peak in the error of $s^2_{23}$. Thus, after $\theta_{13}$ is precisely measured, $\theta_{23}$ is the angle determined with the least accuracy. Its super-precision measurement is not easy unless we can build a dedicated apparatus for CP $\delta$ measurement \cite{Minakata:2013eoa}.
}

Interestingly enough, the bast fit value of $\sin^2 \theta_{23}$ of the T2K new data deviates slightly from the maximal $\theta_{23}$ toward the second octant. It is in agreement with the tendency possessed by the latest analysis of Super-K atmospheric data. See Fig.~16 of \cite{atm-review}. On the other hand, the MINOS best fit is in the first octant of $\theta_{23}$. It is a tantalizing question whether the value of $\sin^2 \theta_{23}$ stays at maximal or deviates from it, and to which way if deviates. 

\subsection{1-2 sector parameters $\Delta m^2_{21}$ and $\theta_{12}$}

\begin{figure}[htb]
\vglue 3mm
\centering
\vglue 3mm
\includegraphics[width=0.48\textwidth]{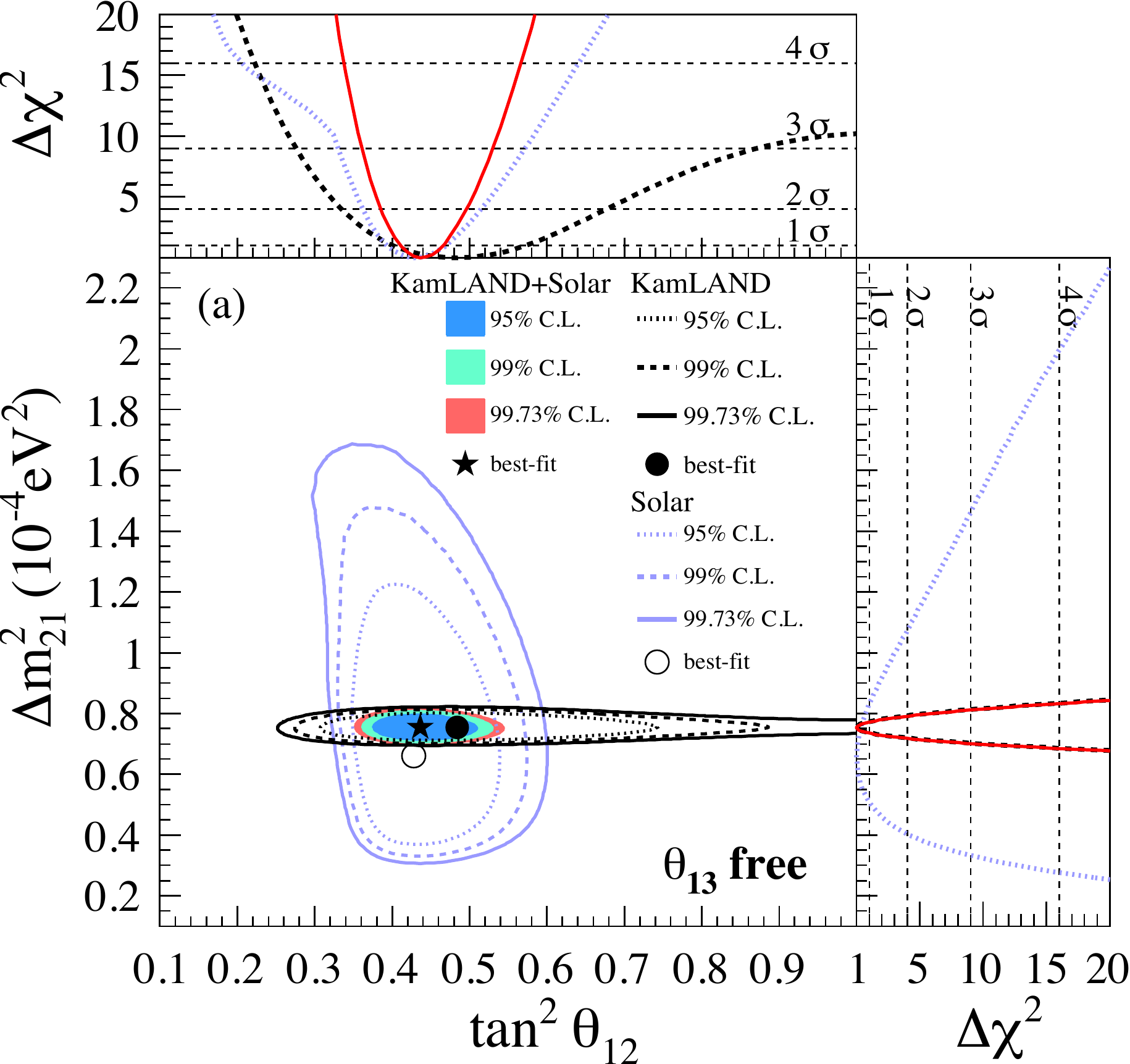}
\includegraphics[width=0.48\textwidth]{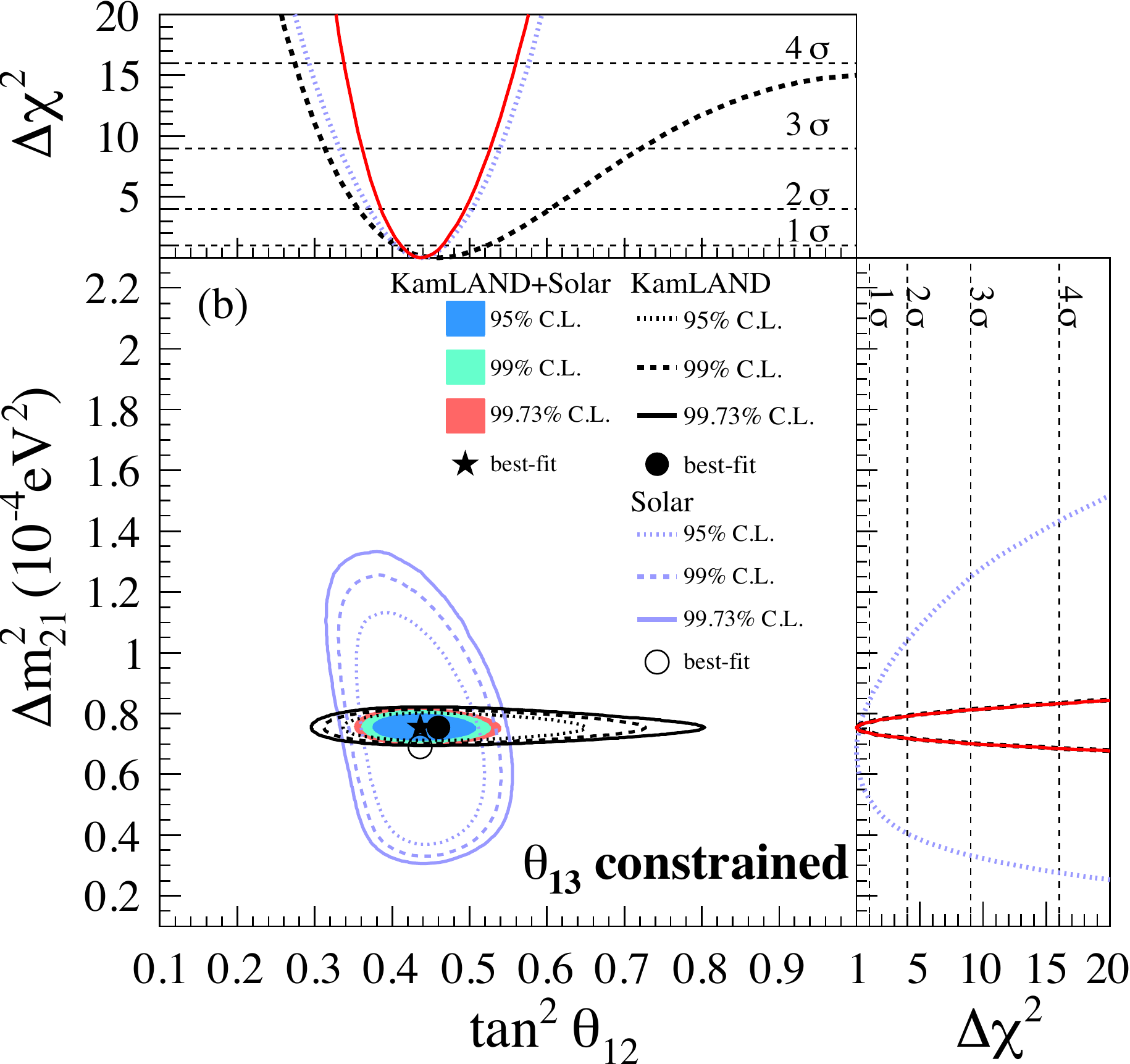}
\caption{
The regions in $\tan^2 \theta_{12} - \Delta m^2_{21}$ space allowed by the KamLAND data (black lines) and all the solar neutrino data combined (blue lines).
The left (right) panel is without (with) $\theta_{13}$ constraint imposed by the short-baseline reactor and accelerator experiments. 
The figures are taken from \cite{Gando:2013nba}.
}
\label{fig:12-parameters}
\end{figure}

The most precisely measured parameters $\Delta m^2_{21}$ and $\theta_{12}$ in the lepton mixing are both in the 1-2 sector. Its current status is shown in Fig.~\ref{fig:12-parameters}. The left (right) panel is without (with) $\theta_{13}$ constraint imposed by the short-baseline reactor and accelerator experiments. As it stands, KamLAND (having reactors with variety of distances in $100-200$ km) surpasses in the accuracy of $\Delta m^2_{21}$, whereas the solar neutrino experiments have better sensitivity to $\theta_{12}$. By combining these two different measurement, the errors of $\Delta m^2_{21}$ and $s^2_{12}$ are only $\simeq 2.4\%$ and $\simeq 4.3\%$ levels, respectively.

%
\begin{figure}[htb]
\vglue 2mm
\centering
\vglue 2mm
\includegraphics[width=0.6\textwidth]{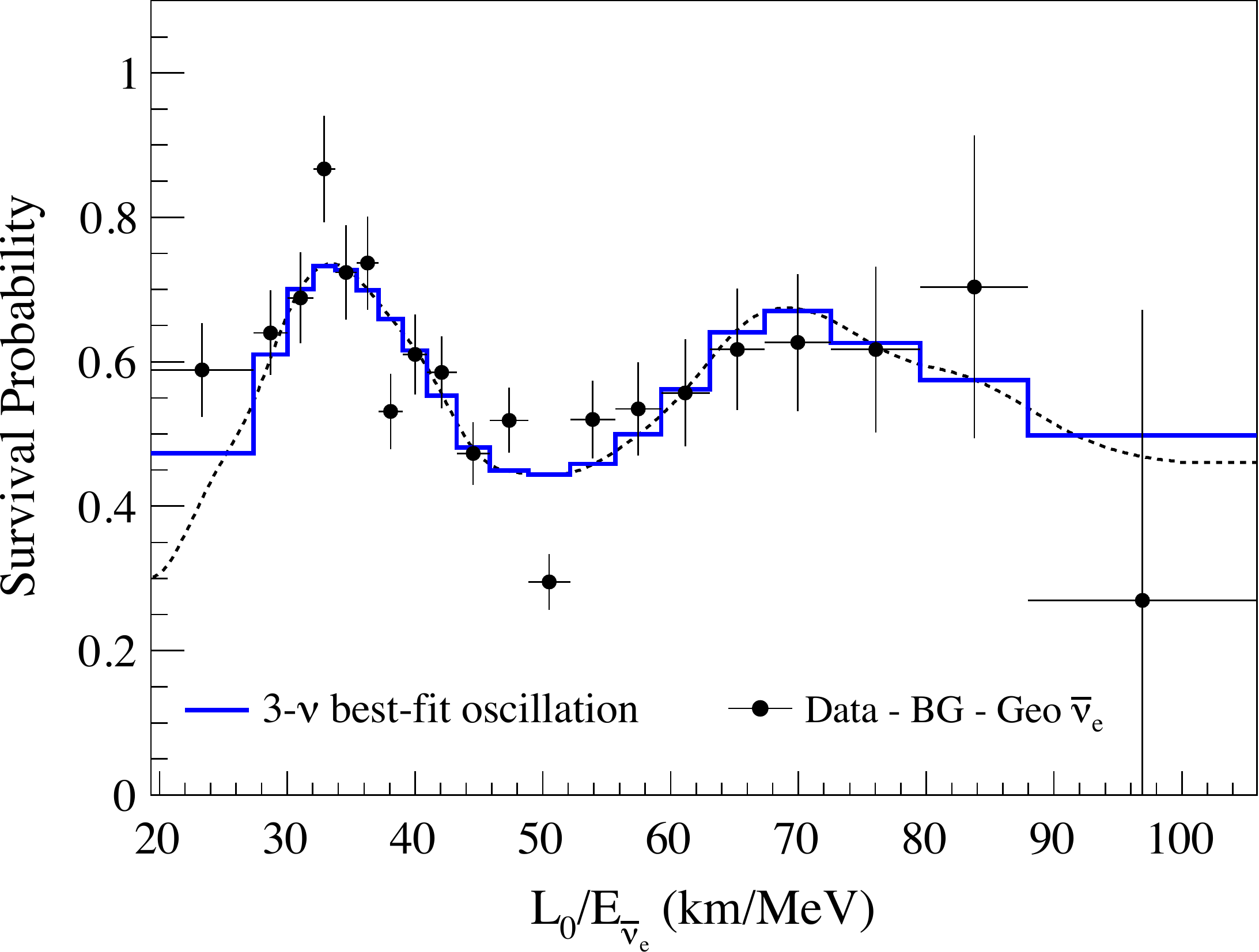}
\caption{
Ratio of the observed $\bar{\nu}_e$ spectrum to the expectation for no-oscillation versus $L_0 / E$ for the KamLAND data. $L_0 = 180$ km is the flux-weighted average reactor baseline. 
This figure is taken from \cite{Gando:2013nba}.
}
\label{fig:KL-oscillation}
\end{figure}

It is interesting to note that the $\theta_{13}$-free fit (left panel) results in the best fit value of $\sin^2 \theta_{13} = 0.023^{+ 0.015}_{-0.015}$. It may be compared to the one of global fit with the reactor $\theta_{13}$ measurement, $\sin^2 \theta_{13} = 0.023^{+ 0.002}_{-0.002}$ \cite{Gando:2013nba}. That is, the solar and the KamLAND experiments by themselves  are now able to pin down the value of $\theta_{13}$, but with much larger errors.
By observing reactor $\bar{\nu}_{e}$ in varying distances clustered around $\sim 180$ km KamLAND also gives the best proof to date of the oscillatory behavior of neutrino disappearance which spans almost two cycles, as shown in Fig.~\ref{fig:KL-oscillation}.

\subsection{1-3 sector mixing angle $\theta_{13}$}

A year after the announcement of seeing $\nu_{e}$ appearance by T2K \cite{T2K} was the year of ``sturm und drang'' of $\theta_{13}$. After the similar indications of ``large $\theta_{13}$'' by MINOS \cite{Adamson:2011qu} and Double Chooz \cite{Abe:2011fz}, the angle is now measured accurately by Daya Bay \cite{Daya-Bay,An:2013zwz} and RENO \cite{Ahn:2012nd,RENO}:
\begin{eqnarray}
\sin^2 2 \theta_{13} &=& 0.090^{+ 0.008}_{-0.009}
\hspace{10mm} ({\rm Daya~Bay}, 1 \sigma) 
\nonumber \\
\sin^2 2 \theta_{13} &=& 0.10^{+ 0.016}_{- 0.016}
\hspace{10mm} ({\rm RENO}, 1 \sigma) 
\label{daya-bay-RENO}
\end{eqnarray}
where the RENO result in (\ref{daya-bay-RENO}) assumes adding the systematic and statistical errors in quadrature. The error in $\sin^2 2 \theta_{13}$ is less than 10\% now. What was surprising to me is that the Daya Bay spectrum measurement leads to an accurate measurement of atmospheric $\Delta m^2$, 
$\Delta m^2_{32} = 2.59^{+ 0.19}_{-0.20} \times 10^{-3} {\rm eV}$, which is worse than MINOS only by a factor of $\simeq 2$. It is really the power of identical multi-detector setup \cite{Mikaelyan,MSYIS}. 
It would be very interesting to watch whether the reactor and the accelerator measurement of $\Delta m^2_{32}$ continues to agree with each other, or finally develops a difference.

\section{Any regularity ?}

It is a tantalizing question whether there exist any regularities hidden in the measured values of the three mixing angles. I must say that the whole bunch of proposals exist in the literature, which are too numerous to cover here. I mention only a few possibilities. See e.g., \cite{Mohapatra:2006gs,Altarelli:2010gt,King:2013eh,Smirnov:2013uba} for more possibilities.

The best studied example, I guess, of the simple parametrization of the MNS matrix is the tri-bimaximal mixing matrix \cite{Harrison:2002er} 
\begin{eqnarray}
U_{TB} =
\left[
\begin{array}{ccc}
\sqrt{ \frac{2}{3} }&  \frac{1}{\sqrt{3} } & 0 \\
-  \frac{1}{\sqrt{6} } &
  \frac{1}{\sqrt{3} } & \frac{1}{\sqrt{2}} \\
  \frac{1}{\sqrt{6} }  &
-  \frac{1}{\sqrt{3} } & \frac{1}{\sqrt{2}} \\
\end{array}
\right]. 
\label{tri-bimaximal}
\end{eqnarray}
It triggered very many works on how one can perturb around (\ref{tri-bimaximal}). What is interesting is the common feature that such perturbation which produces non-zero $\theta_{13}$ generally leads to deviation from the maximal $\theta_{23}$. 
In this context, it is interesting to note that the recent analysis of Super-K atmospheric neutrinos as well as T2K \cite{Abe:2014ugx} tend to favor the second octant solution of $\theta_{23}$. Unfortunately, even if the feature is confirmed, it does not appear to uniquely select out the type of perturbation. 
To my prejudice, however, the expectation to this approach becomes lower after we know that $\theta_{13}$ is ``large'', $\theta_{13} \simeq 9^\circ$, comparable with the Cabibbo angle. 

The related approach is to break $\mu - \tau$ symmetry (see e.g., \cite{Mohapatra:2006gs}) by which the deviation from maximal $\theta_{23}$ is related to $\theta_{13}$. By using the symmetry breaking in the subdominant block with the breaking parameters $a > b$ ($b > a$) for the second (first) octant $\theta_{23}$ and $c=1$ in the notation of \cite{Mohapatra:2006gs}, one obtain 
$D_{23} \equiv 0.5 - \sin^2 \theta_{23} 
= \frac{1}{4} \left( \frac{b+a}{b-a} \right) \sin^2 \theta_{13}$, which is tunable to any values of $\sin^2 \theta_{23}$ within experimental errors including the above one.

More importantly, it triggered a flow of works to address the question of which symmetry is behind the tri-bimaximal mixing, which resulted intensive research on discrete flavor symmetries, in general. There exist extensive references, as you can find in \cite{Altarelli:2010gt,King:2013eh,Smirnov:2013uba}. If the lepton flavor mixing in fact requires a discrete symmetry it is the sign that the generation structure would originate from a geometry of some space or even space-time (which is yet unknown). 

The quark lepton complementarity \cite{QLC,Raidal:2004iw} is started with the empirical observation $\theta_{12} + \theta_{C} \simeq \frac{\pi}{4}$ (for a review see e.g., \cite{Minakata:2005rf}), where $\theta_{C}$ is the Cabibbo angle. When formulated with a natural ansatz of large (bi-maximal) mixing from the neutrino sector it leads to the relation \cite{QLC} (see also \cite{Giunti:2002pp})
\begin{eqnarray}
\sin^2 \theta_{13} \approx \frac{1}{2} \sin^2 \theta_{C},
\label{QLC-relation}
\end{eqnarray}
which agrees well with the experimental data. 

It appears that a bold suggestion is encouraged in our field, the tradition which I now follow. It appears that the recent analysis of Super-K atmospheric neutrinos tend to favor the inverted mass hierarchy though only slightly. It also prefers the second octant solution of $\theta_{23}$ with the best fit value $\sin^2 \theta_{23} \simeq 0.56-0.58$, and the feature seems to be more robust in the case of inverted mass hierarchy.  See Figs.~15 and 16 of \cite{atm-review}. The tendency has been strengthened further more recently by the fact that T2K saw more events than expected \cite{Abe:2013hdq}. In fact, the result of global analysis by the Bari group, which previously favored the first octant solution of $\theta_{23}$ \cite{Fogli:2012ua}, now prefers the second octant solution for the inverted mass hierarchy. See Fig.~2 in \cite{Capozzi:2013csa}. 
If it is the case and choosing the bottom region in Super-K data, $\theta_{23} \simeq 48^\circ - 50^\circ$. Then, one could think of an empirical relation 
\begin{eqnarray}
\theta_{12} + \theta_{13} + \theta_{23} \simeq 33^\circ + 9^\circ + 49^\circ = 91^\circ \simeq \frac{\pi}{2}; 
\label{sum-rule}
\end{eqnarray}
The sum of $2 \theta$'s can form a triangle, which may stimulate a geometrical interpretation.

\section{What is next? No.~1:  Mass hierarchy}

\begin{figure}[htb]
\vglue -1mm
\centering
\vglue -1mm
\includegraphics[width=0.56\textwidth]{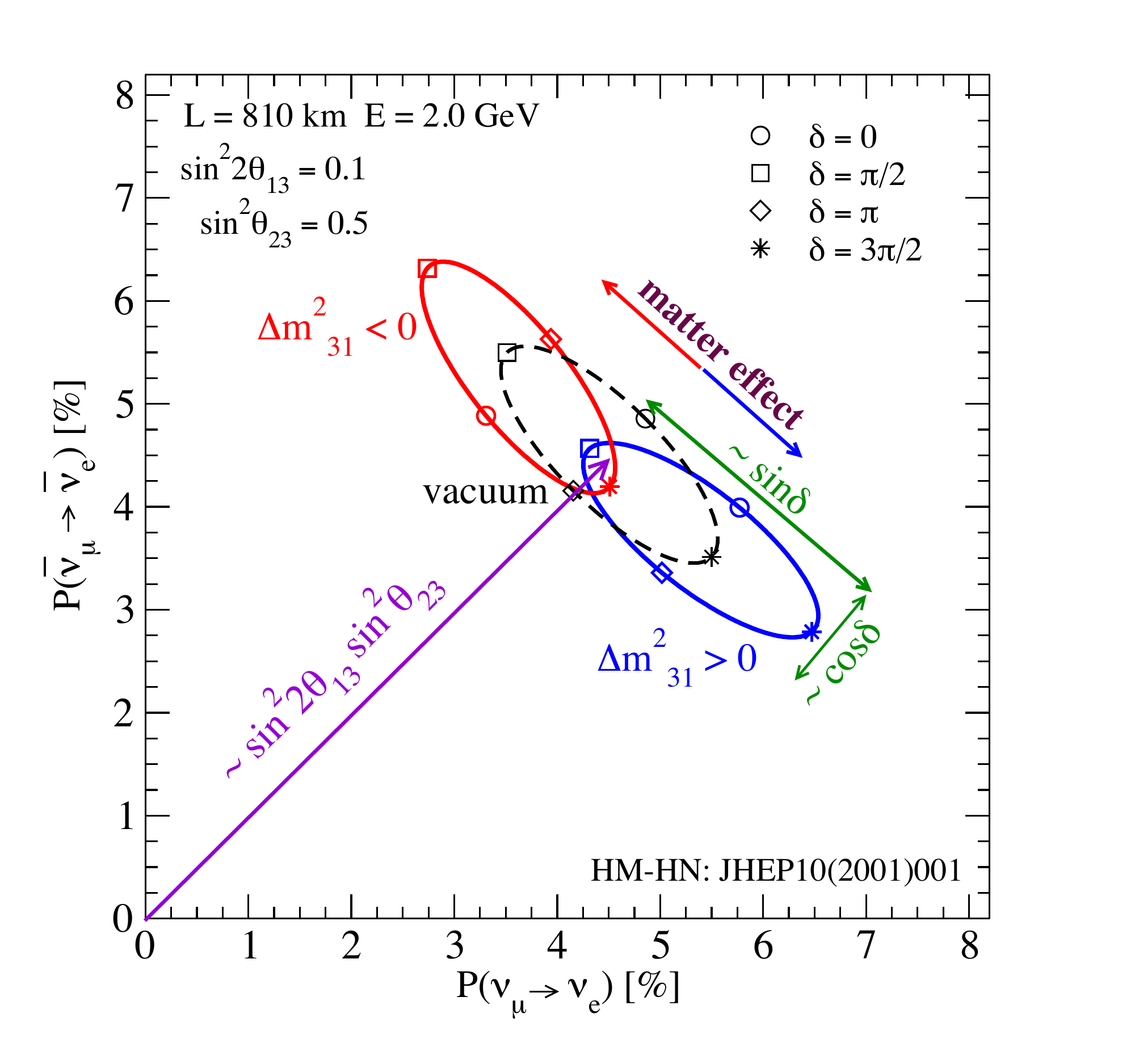}
\caption{
The bi-probability plot in $P(\nu_\mu \rightarrow \nu_e) - P(\bar{\nu}_\mu \rightarrow \bar{\nu}_e)$ space is drawn by taking the baseline $810$ km.
It displays competing three effects, CP violating and CP 
conserving effects due to $\delta$ as well as the matter effects 
in a compact fashion \cite{MN-jhep01}. 
}
\label{fig:bi-P-plot}
\end{figure}

Most probably, the next reachable goal of answering the questions about the unknowns in neutrino physics would be to identify the neutrino mass hierarchy. See e.g., \cite{Minakata:2012ue} for a brief status summary. The principle of determining the mass hierarchy is very simple, and it can be readily understood by using the bi-probability plot in $P(\nu_\mu \rightarrow \nu_e) - P(\bar{\nu}_\mu \rightarrow \bar{\nu}_e)$ space \cite{MN-jhep01}.
When the lepton KM phase $\delta$ is varied the bi-probability trajectory forms an ellipse as shown in Fig.~\ref{fig:bi-P-plot}. The size of the ellipse is a measure for the effect of $\delta$. Whereas the distance between the normal hierarchy ellipse (blue) and the inverted one (red) represents the matter effect. Then, by using the matter effect, which naturally present in the LBL experiments, one can determine the neutrino mass hierarchy. 

The question is then how to realize such experiments. The relative strength between the matter effect to the vacuum effect in neutrino oscillation may be parametrized by the ratio
\begin{eqnarray}
\frac{ a }{ \Delta m^2_{31} } 
=
0.085 
\left(\frac{\rho}{2.8 \ \mathrm{g/cm}^3}\right)
\left(\frac{Y_e}{0.5}\right)
\left(\frac{2.5 \times 10^{-3}\ \mathrm{eV}^2}{\Delta m^2_{31}}\right)
\left(\frac{E}{1 \ \mathrm{ GeV}}\right)
\label{matt-vac-ratio}
\end{eqnarray}
where $a \equiv 2\sqrt{2} G_F N_e E_\nu$ denotes the Wolfenstein matter potential \cite{wolfenstein}. 
On the other hand, the baseline $L$ and the neutrino energy $E$ is related at the vacuum oscillation maximum (VOM) as $\frac{\Delta m^2_{31} L}{4E} = \frac{\pi}{2}$, which leads to $\left(\frac{L}{1000 \ \mathrm{ km }}\right)_{\rm VOM} = 0.495 
\left(\frac{2.5 \times 10^{-3}\ \mathrm{eV}^2}{\Delta m^2_{31}}\right) \\
\times 
\left(\frac{E}{1 \ \mathrm{ GeV}}\right)_{\rm VOM}$. 
Eliminating $E$ factor we obtain, at the first VOM, 
\begin{eqnarray}
\frac{ a }{ \Delta m^2_{31} } = 0.186 
\left(\frac{\rho}{2.8 \ \mathrm{g/cm}^3}\right)
\left(\frac{Y_e}{0.5}\right)
\left(\frac{L}{1000 \ \mathrm{ km }}\right)_{\rm VOM}. 
\label{matt-vac-ratio2}
\end{eqnarray}
Therefore, if one want to remain reasonably close to the vacuum oscillation maximum and at the same time to receive a sizeable matter effect the long baseline $L \sim 1000 - 2000$ km is required. Thus, an intense neutrino beam is necessary to guarantee the event rate sufficient for determining the mass hierarchy. 

The leading candidate for ongoing LBL experiment which has sensitivity to the mass hierarchy is NO$\nu$A in USA \cite{NOVA}. However, the limited statistics may require some luckiness for NO$\nu$A to identify the mass hierarchy. Yet, by taking the large number of $\nu_{e}$ appearance events at T2K \cite{Abe:2013hdq} seriously one may argue that the most likely value of $\delta$ is around $ -\frac{\pi}{2}$, which is the ideal case for NO$\nu$A (and for T2K for CP). If NO$\nu$A alone is not enough to determine the hierarchy the best way is to combine the other experiments, T2K and/or ICAL at INO in India \cite{Blennow:2012gj}. If their sensitivities are not sufficiently high we may need ``dedicated'' LBL experiments. They include LBNE \cite{LBNE}, LBNO \cite{LBNO}, or neutrino factory \cite{nufact} etc.\footnote{
With Kajita-san and the others I put forward the idea of T2KK, Tokai-To-Kamioka-Korea two detector complex which receives an intense neutrino beam from the MW-upgraded JPARC \cite{T2KK}. I still feel it a scientifically viable option. But, it appears that the practical way to realize it is to first construct Hyper-K \cite{HK-LOI} in Japan, and then invite our Korean friends for the Korean detector. 
}

\begin{figure}[htb]
\vglue -2mm
\centering
\includegraphics[width=0.72\textwidth]{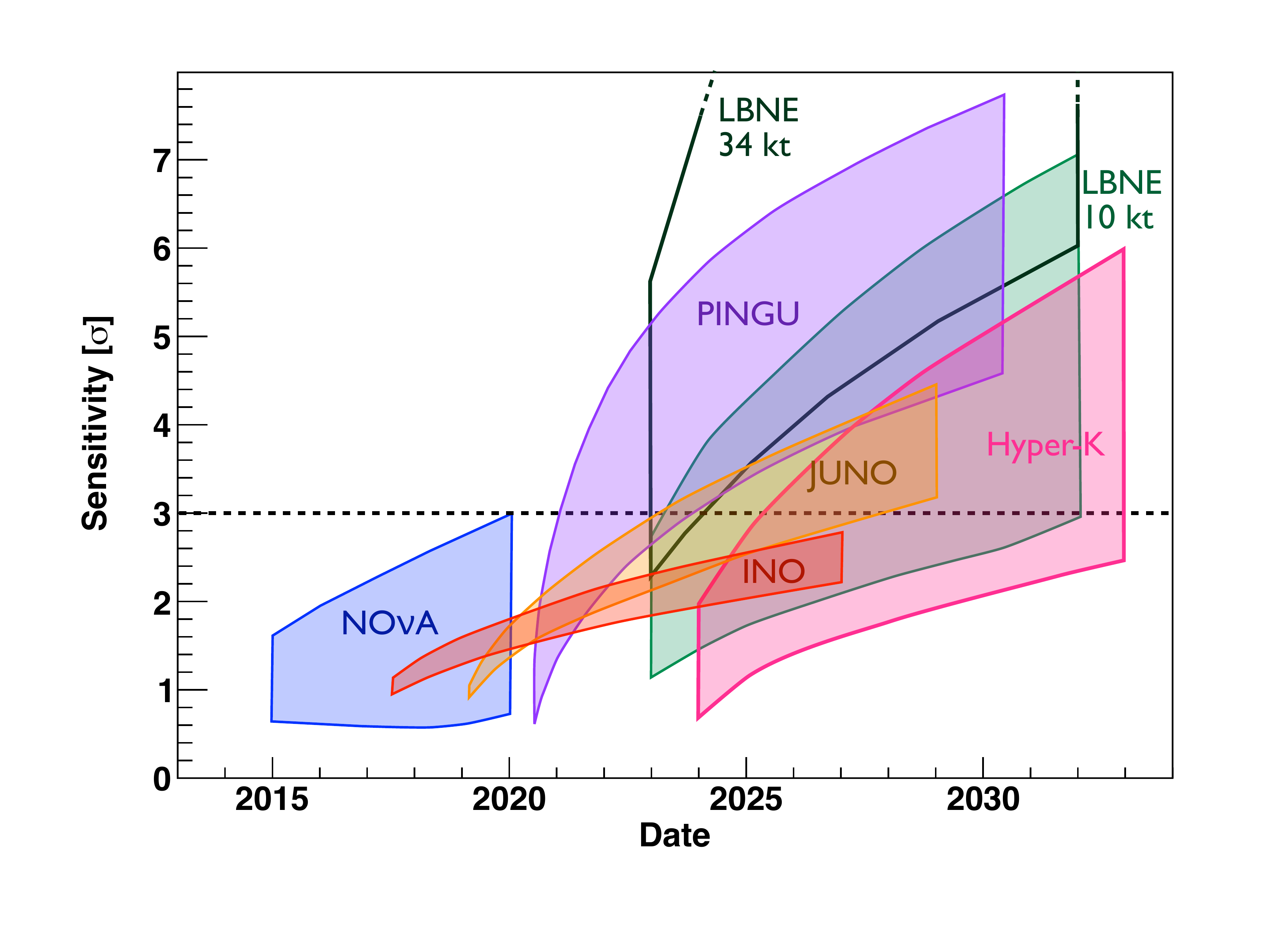}
\vglue -3mm
\caption{
Comparison of the expected sensitivities (for rejecting the inverse hierarchy assuming the normal hierarchy) of different experiments with the potential to measure the neutrino mass hierarchy. The figure is taken from \cite{Aartsen:2014oha} but is kindly simulated by the authors of \cite{Blennow:2013oma}.
}
\label{fig:Pilar}
\end{figure}

There are other ways to utilize the earth matter effect of atmospheric neutrinos which penetrate through the earth before reaching a detector. In this case, we need a large volume detector such as megaton water Cherenkov detector, Hyper-K \cite{HK-LOI}, or high string-density region in IceCube with lowered threshold, PINGU \cite{Aartsen:2014oha}. For earlier exposition for the latter see e.g., \cite{Akhmedov:2012ah}. The relationship between varying apparatus/methods and comparison of capabilities of resolving the mass hierarchy are examined in \cite{Blennow:2013oma}, whose brief summary may be seen in Fig.~\ref{fig:Pilar} (which is actually taken from \cite{Aartsen:2014oha}.)

To explore the matter effect in relationship with mass hierarchy determination any high density environment is in principle adequate. Natural candidates include neutrinos from the sun and from supernovae. For solar neutrino case its low energy $E < 10$ MeV makes neutrino flavor transition in the sun mainly due to solar $\Delta m^2$ effect, which makes the job a bit difficult. For supernova neutrinos the level crossing in the high-density He burning shall is the ideal place to examine the hierarchy issue. (For exposition in the early stage idea see e.g., \cite{Dighe:1999bi,MN-plb01}.) However, there is uncertainty in the flux prediction of neutrinos from supernovae. In the cooling phase, the flux is more or less symmetric with respect to neutrino flavor, preventing a robust determination of the mass hierarchy. Therefore, use of either neutronization burst \cite{Kachelriess:2004ds} or the accreting phase \cite{Buras:2003sn} look more promising. 
The other method is to use the earth matter effect with supernova neutrinos. For detailed examination of this possibility see e.g., \cite{Dighe-Raffelt}, and \cite{Borriello:2012zc} for a recent critical examination.

I must warn the readers that use of the matter effect is {\em not} the only way to resolve the neutrino mass hierarchy. It can be shown on general ground that neutrino oscillation wave in vacuum distinguishes the mass hierarchies in near the maxima of the solar $\Delta m^2$ driven long-wavelength oscillation and only in there \cite{MNPZ2}. This phenomena was used by the authors of \cite{petcov} for a bold suggestion of using reactor neutrinos to resolve the neutrino mass hierarchy. Then, the possibility was seriously taken by some experimentalists \cite{hanohano,Zhan:2009rs}. There exist at least two concrete proposals for such measurement, JUNO in China \cite{Kettell:2013eos} and RENO-50 in Korea \cite{Seo:2013swa}. Since there exist highly nontrivial requirements for the energy resolution and linearity of reconstructing neutrino energy \cite{Qian:2012xh}, the feasibility of this method for determining the mass hierarchy is currently under debate. Therefore, our community is eagerly waiting to see the outcome from such experimental endeavor.  

\section{What is next? No.~2: CP $\delta$ and $\theta_{23}$}
\label{sec:CP}

The principle of determining the lepton KM phase $\delta$ is also very simple. Measurement of neutrino and anti-neutrino appearance oscillation probabilities, $P(\nu_\mu \rightarrow \nu_e)$ and $P(\bar{\nu}_\mu \rightarrow \bar{\nu}_e)$, would determine $\delta$ and $\theta_{23}$ (given accurate measurement of $\theta_{13}$) up to the degeneracies, as can be seen in Fig.~\ref{fig:bi-P-plot}. The degeneracies just mentioned is not limited to the well known intrinsic degeneracy \cite{BurguetCastell:2001ez} and the sign-$\Delta m^2$ degeneracy \cite{MN-jhep01},\footnote{
For a global overview of the structure of the degeneracy excluding generalized intrinsic degeneracy, see e.g., \cite{Minakata:2010zn}.
}
but also include more general structure called ``generalized intrinsic degeneracy'' \cite{CMP} which includes the $\theta_{23}$ intrinsic degeneracy discussed in \cite{Minakata:2013eoa} as a part of it.

Nonetheless, its determination would be the farthest-reaching goal among measurement of all the three-flavor neutrino mixing parameters.\footnote{
The discussion here a priori excludes the important issues of absolute neutrino masses and the Majorana phases, the latter assuming that neutrinos are Majorana particles. See Sec.~\ref{sec:cosmo-particle}.
}
It was particularly nice that $\theta_{13}$ is large, just below the Chooz-Palo-Verde bound. Yet, by being the genuine three-flavor effect, it is suppressed by the two small factors, the small ratio $\frac{\Delta m^2_{21}}{\Delta m^2_{31}} \simeq 0.031$ and the reduced  Jarlskog factor $J_r = c_{12} s_{12} c_{23} s_{23} c^2_{13} s_{13} \simeq 0.035$.\footnote{
Notice that the statement applies not only to CP violating $\sin \delta$ effect but also (after a minor correction) to CP conserving $\cos \delta$ effect, as generally proved in \cite{Asano:2011nj}. The minor correction needed is the missing $c^2_{13}$ factor so that $\cos \delta$ effect does not vanish at $\theta_{13} = \frac{\pi}{2}$.
}
Therefore, the effect is typically of the order of $\sim 10^{-3}$. Thus, the bottom line is that we need a dedicated machine for CP to measure the lepton KM phase $\delta$. 

Now, there are several ideas or proposals to realize measurement of CP violating phase. They include: Hyper-K \cite{HK-LOI}, LBNE \cite{LBNE}, LBNO \cite{LBNO}, and neutrino factory \cite{nufact} etc. The proposals are all well thought ones and they will certainly be able to measure the effect of $\delta$. A potential problem is that they are all costly and requires $\sim 10$ years to construct. Of course, if one of them declares its construction next month, it would be great!

But, if it does not happen soon, what shall we do? What we can do is to combine existing measurements to extract information of $\delta$. Conventionally, it is carried out by exploiting the strategy called the ``global analysis'' by putting everything in it. The well known groups which engage this painful task include the authors of Refs.~\cite{Tortola:2012te,Fogli:2012ua,GonzalezGarcia:2012sz}. Of course, there is nothing wrong with it. 
%
We took a different attitude by combing not everything but only the relevant ones, the short-baseline reactor and accelerator appearance measurement for $\theta_{13}$, for example \cite{Machado:2011ar}. There are pros and cons in each approach. In the global fit the sensitivity is higher, but it is achieved at the price of combining many experiments with different systematic errors. In our approach that drawback is somewhat cured though it may not reveal the best possible sensitivity to CP violation. I believe that it is important to proceed with the two approaches which are complementary to each other.

For CP phase $\delta$ we continued with our way of thinking \cite{Machado:2013kya}. For the closely related works see e.g., \cite{Huber:2009cw,Ghosh:2013yon}. For the time being the CP sensitivity that can be achieved by ongoing and upcoming LBL neutrino oscillation experiments is quite limited. To know the state of the art we have combined $\sim 10$ years running of T2K and NO$\nu$A under the constraint of short-baseline reactor $\theta_{13}$ measurement \cite{Minakata:2003wq}. 
In \cite{Machado:2013kya} we have used ``CP exclusion fraction'' $f_{\rm CPX}$, the fraction of CP values of $\delta$ which can be disfavored by these experiments for a given set of input parameters.\footnote{
The CP exclusion fraction plot is a particularly useful tool to reveal the potential for exploring the CP phase effects by a ``non-conclusive experiment'' which is not designed as a dedicated CP violation discoverer. 
Suppose that there are two experiments each of which alone can not discover (establish) CPV at a given CL. In this case the CPV fraction vanishes for both experiments, and it does not provide us with any useful informations. But, with use of $f_{\rm CPX}$ we are able to reveal CP sensitivity of each experiment and can tell which one has higher capability of restricting the allowed range of $\delta$. In this way, the CP exclusion fraction serves as a viable way of quantifying the experimental CP sensitivity for non-conclusive experiments, and provides a better chance for fruitful discussion of synergy. $f_{\rm CPX}$ is closely related with the measure proposed earlier \cite{Winter:2003ye}.
}
It is thus a global measure which covers an entire input parameter space, as is the case of popularly used CP fraction.

\begin{figure}[htb]
\vglue -2mm
\centering
\includegraphics[width=0.24\textwidth]{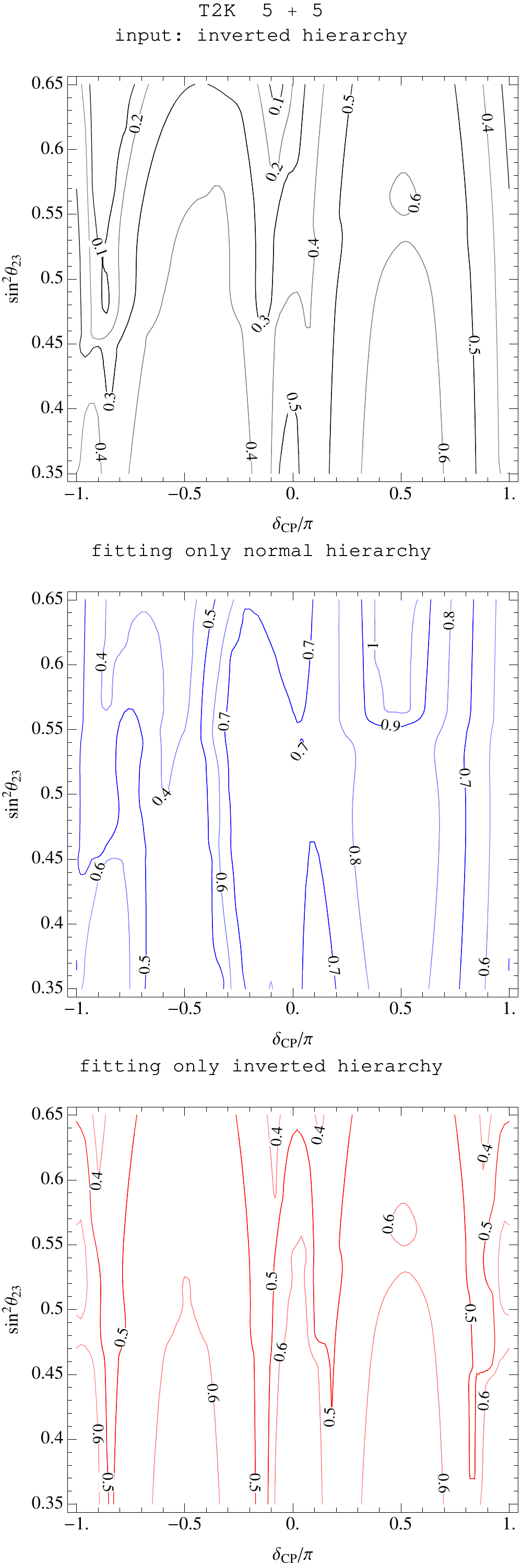}
\includegraphics[width=0.24\textwidth]{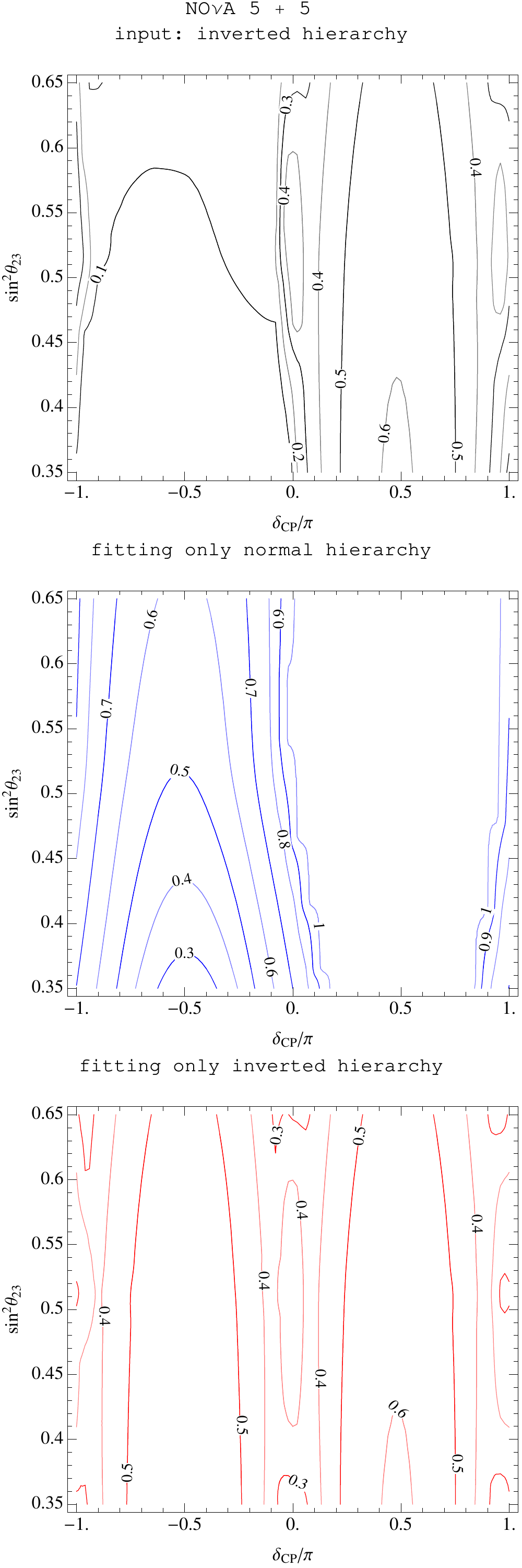}
\includegraphics[width=0.24\textwidth]{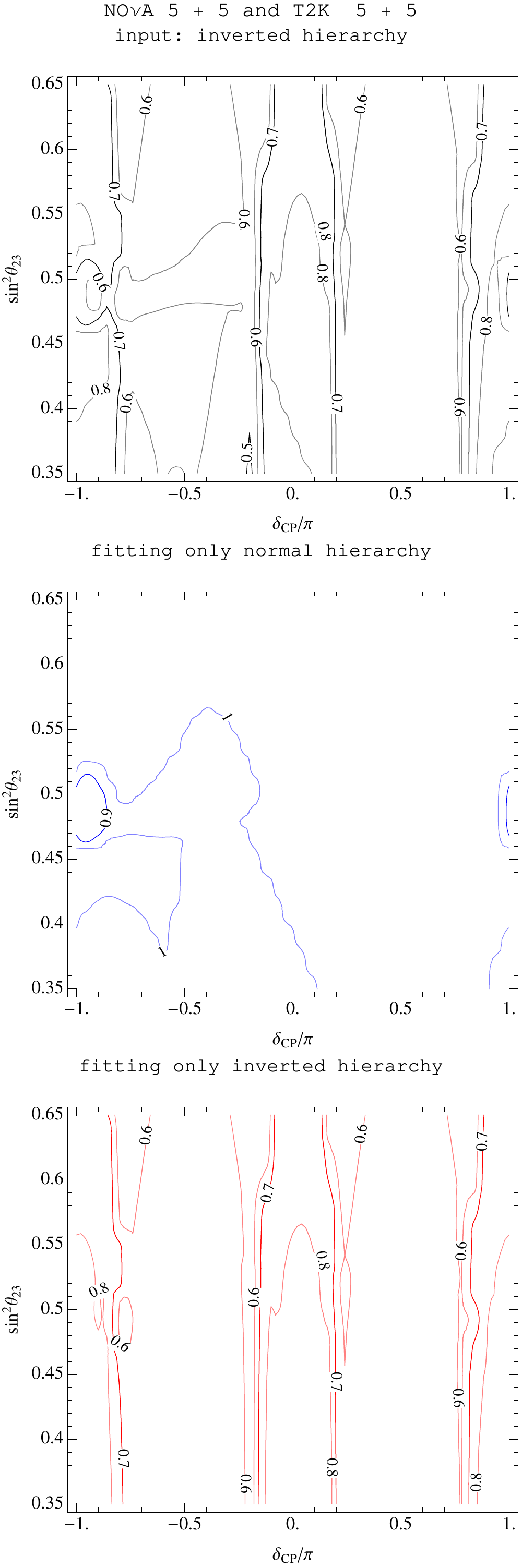}
\vglue -2mm
\caption{
Presented are the iso-contours of $f_{\rm CPX}$ on the $\delta - \sin^2 \theta_{23}$ ($\delta_{\rm CP} = \delta$ in our notation) plane at 90 \% CL  assuming the input (true) inverted mass hierarchy. The left and middle panels are the results of 5 years running with $\nu$ and 5 years running with $\bar \nu$ modes for T2K and NO$\nu$A, respectively. Whereas the right panels display the result of combining $5+5$ years running of T2K and NO$\nu$A. 
From top to bottom panels, we marginalize over the hierarchies, fit by assuming the normal mass hierarchy, and fit by assuming the inverted mass hierarchy. 
}
\label{fig:T2K+NOVA}
\end{figure}

In Fig.~\ref{fig:T2K+NOVA}, presented are the iso-contours of $f_{\rm CPX}$ on the $\delta - \sin^2 \theta_{23}$ ($\delta_{\rm CP} = \delta$ in our notation) plane at 90 \% CL  assuming the input (true) inverted mass hierarchy. The left and middle panels are the results of 5 years running with $\nu$ and 5 years running with $\bar \nu$ modes for T2K and NO$\nu$A, respectively. On the other hand, the right panels display the result of combining $5+5$ years running of T2K and NO$\nu$A. From top to bottom panels, we marginalize over the hierarchies, fit by assuming the normal mass hierarchy, and fit by assuming the inverted mass hierarchy. 

General tendency in Fig.~\ref{fig:T2K+NOVA} is that T2K has higher sensitivity to $\delta$, but NO$\nu$A has higher power of rejecting the wrong hierarchy because of its longer baseline. For details of our setup of T2K and NO$\nu$A, and the features with the input normal mass hierarchy, see \cite{Machado:2013kya}. 
As you see the ongoing experiments, under a ``bold assumption'' of running 10 years, the reachable sensitivity to CP phase is not that small. If the mass hierarchy is known, T2K and NO$\nu$A alone may exclude, respectively, about $50\%-60\%$ and $40\%-50\%$ of the $\delta$ space at 90\% CL by 10 years running, provided that a considerable fraction of beam time is devoted to the antineutrino run.  
When T2K and NO$\nu$A is combined most of the region has values of $f_{\rm CPX}$ higher than 0.9 in the wrong hierarchy fit. Thus, the synergy between T2K and NO$\nu$A is remarkable. 

Now, we turn to the measurement of $\delta$ in the dedicated apparatus such as Hyper-K or LBNE. Is everything well understood there at least at the theoretical level? The answer appears to be {\em No}. We have recently addressed a new way of setup of the problem, a simultaneous determination of $\sin^2 \theta_{23}$ and $\delta$ \cite{Minakata:2013eoa}. We are originally motivated by the fact that the prevailing error of $\sin^2 \theta_{23}$ (as discussed in Sec.~\ref{sec:2-3}) is a major limiting factor to precision measurement of $\delta$. Of course, it is a natural and inevitable setting, given the precision measurement of $\sin^2 \theta_{13}$, whose accuracy would eventually reach $\simeq 5\%$ thanks to cancellation of systematic errors due to identical multi-detector setting. In \cite{Minakata:2013eoa} it was shown that when the dedicated experiment starts to measure $\sin \delta$ with uncertainty $\Delta (\sin \delta)$ precision of $\sin^2 \theta_{23}$ is guaranteed to be
\begin{eqnarray}
\Delta (s^2_{23}) \simeq \frac{1}{6} \Delta (\sin \delta), 
\label{error-relation}
\end{eqnarray}
near the first VOM. Therefore, $\nu_e$ appearance measurement is the key to precision measurement of $\sin^2 \theta_{23}$.

Finally in this section I briefly mention CP phase measurement at the second VOM, $\frac{\Delta m^2_{31} L }{ 4 E } = \frac{3 \pi}{2}$. The $\delta$ dependent term in the oscillation probability $P(\nu_\mu \rightarrow \nu_e)$ or $P(\bar{\nu}_\mu \rightarrow \bar{\nu}_e)$ is an interference term between the dominant atmospheric $\Delta m^2_{31}$ scale oscillation term and the solar $\Delta m^2_{21}$ scale oscillation term. At the second VOM the ratio of $\delta$ dependent term to the dominant atmospheric term is three times larger than the one at the first VOM. (Notice that the Korean detector in the T2KK setup \cite{T2KK} is based on the same idea.) Of course, given a beam line one has to go to longer distance by a factor of three, leading to a factor $\sim 10$ fewer number of events. Therefore, CP phase measurement at the second VOM is, in principle, advantageous, if one can overcome the loss of statistics at the longer distance. Such idea has recently embodied by a concrete proposal of ESS$\nu$SB in Europe~\cite{Baussan:2013zcy}. 

The all above discussed will help us to finally win the long-term race, perhaps the hardest one, of hunting the lepton CP phase, {\em the marathon in neutrino physics}.

\section{Dawn of high energy neutrino astronomy}

\begin{figure}[htb]
\centering
\vglue 1mm
\includegraphics[width=0.8\textwidth]{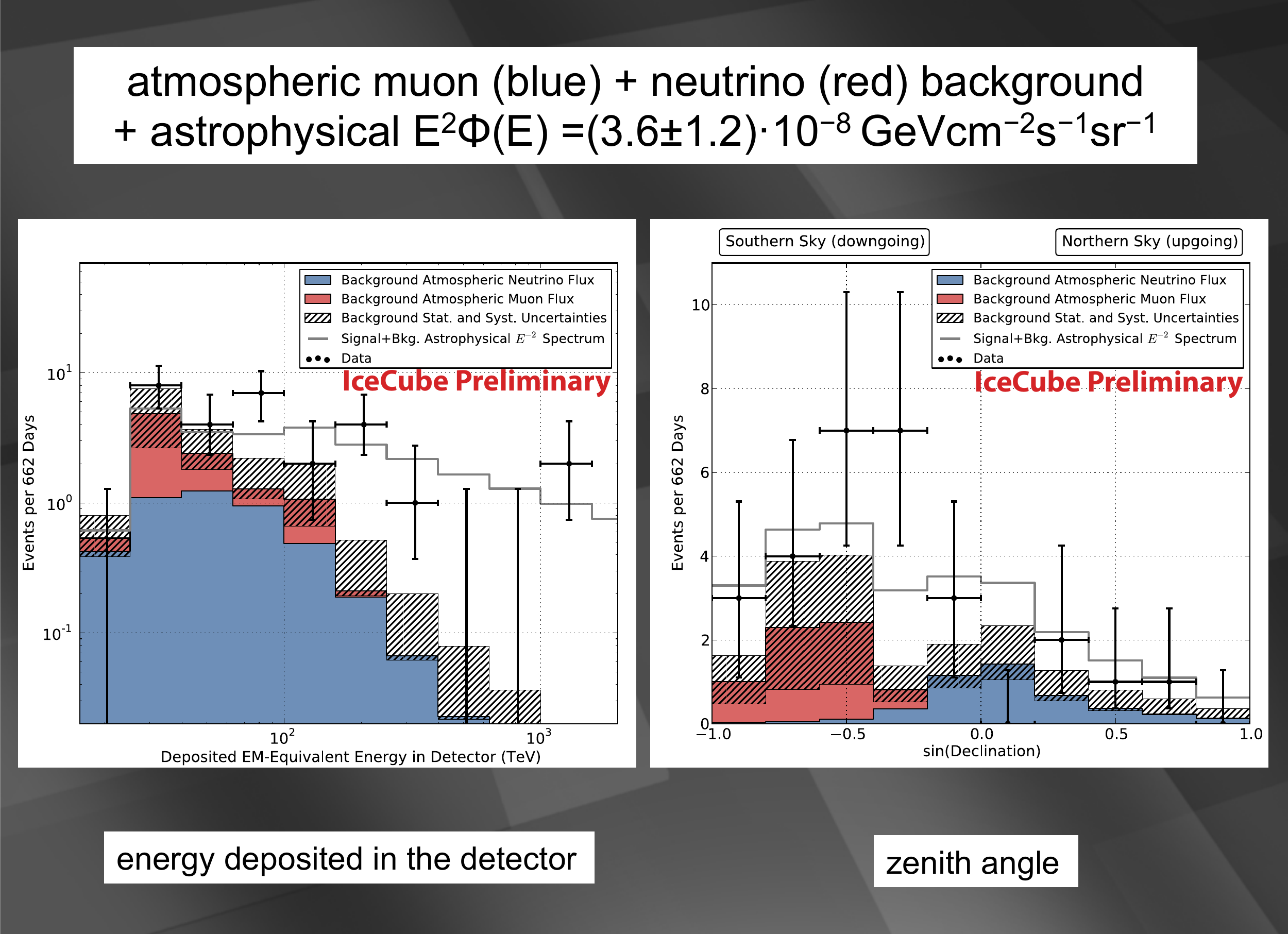}
\vglue 1mm
\caption{
Distribution of the deposited energies (left panel) and declination angles (right panel) of the observed events compared to model predictions.
This figure is taken from \cite{Halzen:2013dva}. 
}
\label{fig:icecube-28}
\end{figure}

Now, let me turn to a completely new direction in this and the next sections, that is, astrophysical and cosmological neutrinos. In 2013 we have observed the dawn of high energy neutrino astronomy.\footnote{
The low energy neutrino astronomy was pioneered by Davis and Koshiba by detecting neutrinos which comes from the sun and the supernova \cite{Nobel-2002}. The high-energy counterpart that has just born will reveal, at its minimum, the secret of high energy universe such as physics of GRBs, AGN, pulsars, and SN shock acceleration, etc.
}
IceCube saw the two $\sim$PeV events which are far above expectation of atmospheric neutrinos \cite{Aartsen:2013bka}. Then, a new analysis which focus on contained showers adds 26 more events at relatively lower energies, altogether makes it to 28 events, the evidence for the excess at $\sim 4 \sigma$ CL above the atmospheric background of about 10 \cite{Aartsen:2013jdh}. See Fig.~\ref{fig:icecube-28} for the energy (left panel) and declination angle distributions. 
It appears that the lack of muon track in most (21) of the events makes it very difficult to believe that the events come from high energy atmospheric neutrinos. The large attenuation effect inside the earth matter makes the down-going events more numerous than up-going ones. 

The origin of such high energy probably extra-galactic (or galactic) neutrinos are much debated in the literature, among which only a very small subset is cited here. For more references see e.g., \cite{Halzen:2013dva}. It appears that not only the atmospheric neutrinos but also the cosmogenic neutrinos are disfavoured as a source of the PeV events \cite{Laha:2013lka}. It is discussed that a power-law neutrino spectrum with index $\Gamma \sim 2.3$ is consistent with the data up to 2 PeV \cite{Anchordoqui:2013qsi}. On the other hand, a gap between lower energy cluster of events and the two PeV events naturally triggered ``new physics interpretation'' such as due to decay of superheavy dark matter scenario \cite{Feldstein:2013kka,Esmaili:2013gha}. The issue of identical or different nature of PeV and lower energy events will be better understood by the coming higher statistics IceCube data in the future. 

\section{New era of cosmology and particle physics}
\label{sec:cosmo-particle}

The year 2013 was an epoch making year in cosmology in which the Planck satellite reported the results of their observation in the initial 16 months since ranched in 2009 \cite{Ade:2013zuv}, bringing cosmology to a truly precision measurement science. Furthermore, it appears that some features seen in the Planck data set triggers renewed interests in learning about neutrinos by the cosmological observations. For earlier references see e.g., \cite{Abazajian:2011dt} and the references cited therein. First of all, the Planck group placed the severe bound on sum of the neutrino masses $\Sigma \equiv \sum_{i = 1,2,3} m_i$ \cite{Ade:2013zuv} 
\begin{eqnarray}
\Sigma < 0.23~{\rm eV} 
\hspace{4mm}
(95\%~CL; {\rm Planck+WMAP-pol+highL+BAO}).
\label{mass-bound}
\end{eqnarray}
Furthermore, the coming precision measurements of galaxy correlation and weak lensing are expected to tighten up the accuracies of $\Sigma$ determination to a level of $\Sigma = 0.05 - 0.02$ eV \cite{Kitching:2008dp,Carbone:2010ik,Amendola:2012ys,Hamann:2012fe}. 

\begin{figure}[htb]
\vglue 2mm
\centering
\includegraphics[width=0.50\textwidth]{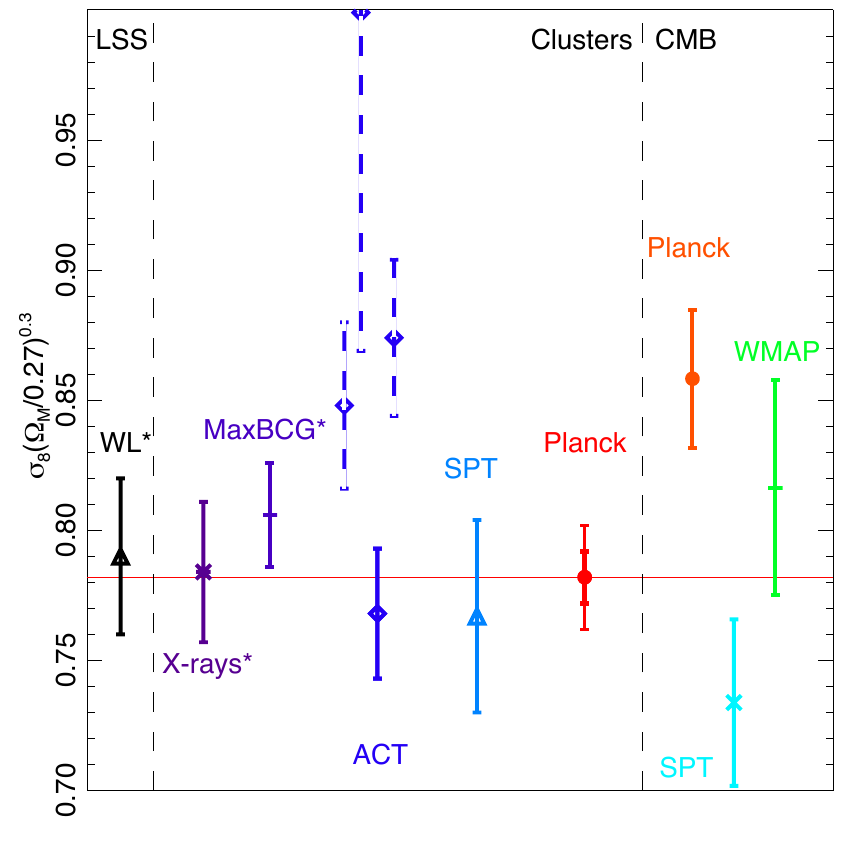}
\caption{
Comparison of constraints ($1 \sigma$ CL interval) on $\sigma_8 (\Omega_\mathrm{m}/0.27)^{0.3}$ from different experiments, taken from \cite{Ade:2013lmv}. They include large--scale structure (LSS), clusters, and CMB. For details see \cite{Ade:2013lmv}.
  }
\label{fig:sigma8}
\end{figure}

Interestingly enough, an apparent discrepancy (see Fig.~\ref{fig:sigma8}) between the matter density correlation parameter at 8 Mpc, $\sigma_{8}$, that deduced from CMB and the late-time observables such as Sunyaev-Zeldovich cluster counts \cite{Ade:2013lmv} and lensing observations \cite{Ade:2013tyw,vanEngelen:2012va,Kilbinger:2012qz} stimulates explanation by the role of active as well as sterile neutrinos \cite{Wyman:2013lza,Wyman:2013lza,Hamann:2013iba}. It would be very interesting to see the outcome of such analyses in the light of future data release from the Planck group. 

Then, what are the implications of the results of these precision cosmological observations? A fuller answer to this question is far above my ability one can certainly say that we are facing with an entirely new era in which the tie between particle physics and cosmology has never been so strong. At least from the point of view of neutrino physics it seems to be the immediate reality. 

\begin{figure}[htb]
\vglue -2mm
\centering
\includegraphics[width=0.66\textwidth]{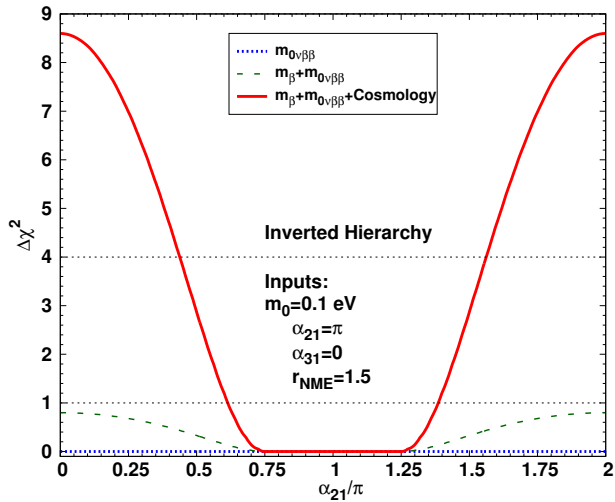}
\vglue -2.6mm
\caption{
$\Delta \chi^2$ is plotted as a function of the fitted value 
of $\alpha_{21}$ for the case of the inverted (left panel) 
and normal (right panel) mass hierarchies. 
The true values of the parameters are taken as $m_0 = 0.1$ eV, 
$\alpha_{21} = \pi$, and $\alpha_{31} = 0$. 
The three $\Delta \chi^2$ curves are presented which correspond to three
 combinations of the data used in the analysis, only $0\nu\beta\beta$
 decay (dotted blue line), $0\nu\beta\beta$ + $\beta$ decays (dashed
 green curve) and all combined, $0\nu\beta\beta$ + $\beta$ decays +
 cosmology (solid red curve). 
}
\label{fig:Mphase-principle}
\end{figure}
%
\begin{figure}[htb]
\vglue -4mm
\centering
\includegraphics[width=0.82\textwidth]{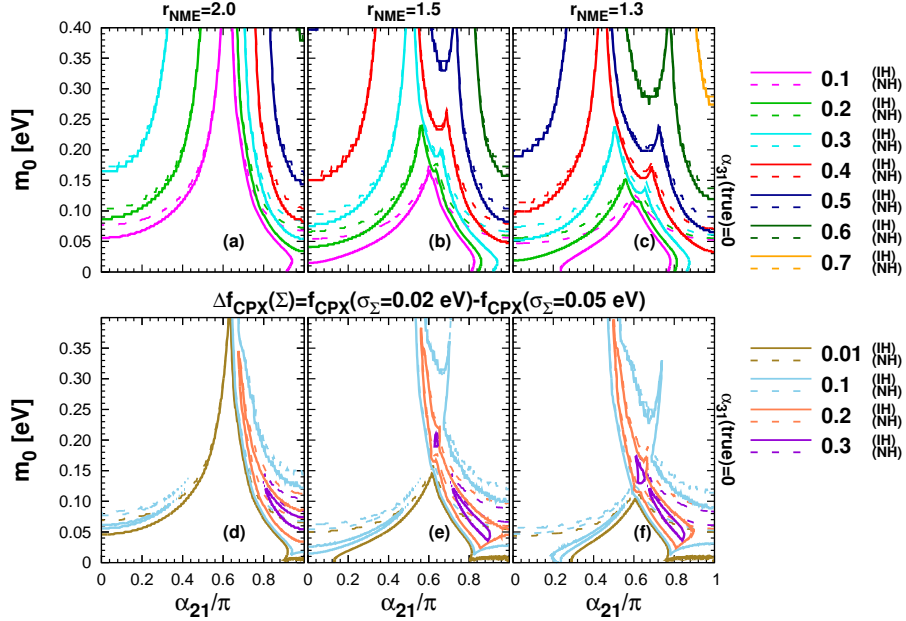}
\vglue -3mm
\caption{
In the upper three panels, the iso-contours of CP exclusion fraction $f_{\rm CPX}$ determined at 2 $\sigma$ CL (1 DOF) with the assumed 1$\sigma$ error of $\Sigma \equiv \sum_{i} m_i$, $\sigma_{\Sigma}=0.02$ eV, are presented to indicate the sensitivity to the Majorana phase $\alpha_{21}$. 
In the lower three panels, the improvement of the sensitivity that was achieved when the error of $\Sigma$ went down from $\sigma_{\Sigma}=0.05$ eV to $\sigma_{\Sigma}=0.02$ eV are indicated by showing $\Delta f_{\rm CPX} (\Sigma) \equiv  f_{\rm CPX}(\Sigma = 0.02 {\rm ~eV}) 
- f_{\rm CPX}(\Sigma = 0.05 {\rm ~eV})$. 
The left, middle, and the right panels are with the uncertainty of nuclear matrix elements of factor $r_{\rm {\tiny NME}} = 2$, 1.5, and 1.3, respectively.  
 The case for the inverted (IH) 
 and normal (NH) mass hierarchy are 
 shown, respectively, by the solid and dashed curves, 
 from 0.1 to 0.7 with the step size of 0.1. The figure is from 
 \cite{MNQ-Majorana-phase}.
 }
\label{fig:f_CPX-majorana}
\end{figure}

Recently, we have analyzed in the light of precision measurement of $\Sigma$ the sensitivity to the Majorana phase (mostly one of them) achievable by the next generation neutrinoless ($0 \nu$) double beta decay experiments \cite{MNQ-Majorana-phase}. (For the foregoing analyses see references in \cite{MNQ-Majorana-phase}.) Needless to say, the nature of neutrinos, Mojorana or Dirac particles, is one of the most important questions among the others in particle physics.
Though the phase is expected to has intimate relationship with the leptogenesis scenario \cite{leptogenesis} which explains baryon number asymmetry in the universe there was prevailing pessimism in the past about capability to detect its effect. Fortunately, we start to enjoy the precision era of cosmological observation, as I emphasized above. The precision expected to the future measurement of $\Sigma$ is sufficiently accurate to constrain or even measure the Majorana phase. 

The principle of constraining the Majorana phase is in fact very simple but not so trivial. First of all, a $0 \nu$ double beta decay experiment measures effective neutrino mass given by  
\begin{eqnarray}
m_{0\nu\beta\beta} 
=
\biggl | m_1 c_{13}^2 c_{12}^2 + 
m_2 c_{13}^2 s_{12}^2 e^{i\alpha_{21}} + 
m_3 s_{13}^2 e^{i\alpha_{31}}  \biggr |.
\label{eq:0nubb-obs}
\end{eqnarray}
Then, since they are sensitive to the Majorana phase and absolute neutrino mass scale they themselves can determine or constrain the Majorana phase in the absence of the uncertainty of nuclear matrix elements. But, it is {\em not} the case. As exhibited in Fig.~\ref{fig:Mphase-principle}, even under the rather optimistic assumption of the small error of $m_{0\nu\beta\beta}$, $\sigma_{0\nu\beta\beta} = 0.01$ eV, essentially no sensitivity to $\alpha_{21}$ can be obtained \cite{MNQ-Majorana-phase}. Notice that the flat bottom at around $\alpha_{21}=\pi$ (assumed true value) is due to the uncertainty of nuclear matrix elements, but the lack of the sensitivity to the phase $\alpha_{21}$ occurs outside the flat bottom. Now, what happens if additional information is provided by either the cosmological observation or single beta decay experiment KATRIN \cite{Wolf:2008hf}. By comparing the red solid curve ($\Sigma$ measurement  $+m_{0\nu\beta\beta}$) to the blue dotted one (only $m_{0\nu\beta\beta}$) in Fig.~\ref{fig:Mphase-principle} independent measurement of neutrino mass scale is the {\em key} to the sensitivity to $\alpha_{21}$. Or, in other words, practically any measured result of $m_{0\nu\beta\beta}$ can be reconciled with any values of the Majorana phase by adjusting the absolute scale of neutrino masses.

To display the sensitivity to the Majorana phase $\alpha_{21}$ in a quantitative way, we have used the CP exclusion fraction $f_{\rm CPX}$, a global measure for the sensitivity to CP violating phase, that has been utilized in our analysis on the lepton KM phase $\delta$. See Sec.~\ref{sec:CP} for the definition of $f_{\rm CPX}$. In Fig.~\ref{fig:f_CPX-majorana}, the $2\sigma$ iso-contours of $f_{\rm CPX}$ is presented in space spanned by $\alpha_{21}/\pi$ and the lightest neutrino mass $m_0$. $m_0=m_1$ for the normal, and $m_0=m_3$ for the inverted mass hierarchies, respectively. 
As can be seen in Fig.~\ref{fig:f_CPX-majorana}, under a rather optimistic assumption of error of $\Sigma$ equal to 0.02 eV and $\sigma_{0\nu\beta\beta} = 0.01$ eV, one can exclude $10\% - 50\%$ of the $\alpha_{21}$ phase space at $2 \sigma$ CL for $m_0 = 0.1$ eV. I think it a great news, but I should remind you that it would only became possible thanks to the expected precision measurement of $\Sigma$ and $m_{0\nu\beta\beta}$. 

However, we note that improvement of sensitivity to $\alpha_{21}$ achieved by decreasing the error of $m_{0\nu\beta\beta}$ from 0.05 eV to 0.02 eV is rather modest, as seen in the lower panels of Fig.~\ref{fig:f_CPX-majorana}. Also, improvement that can be made by the painful effort of improving the uncertainty of nuclear matrix elements from $r_{\rm {\tiny NME}} =1.5$ to 1.3 is also modest, indicating the basic difficulty of the task of measuring the Majorana phase. 
For more about the details in the analysis and the results see \cite{MNQ-Majorana-phase}. 

When the new era of cosmology and particle physics blossoms I would expect that more surprises will be waiting for us before the next CosPA meeting happens.

\section{Final speculation}

I would like to conclude with ``Final speculation'' rather than ``Final remarks''.
Now we see accumulation of very interesting experimental features which trigger optimism, though tantalizing ones because of their low CL, mostly less than $2 \sigma$. They are:

\noindent
(1) The T2K best fit values of $\delta$ is at around $- \frac{\pi}{2}$, which means the best case for NO$\nu$A; It would allow them to determine the mass hierarchy. Then, the detection of CP violation would be in reach in shorter perspective than expected.

\noindent
(2) The Super-K atmospheric neutrino analysis prefers the inverted mass hierarchy, and the second octant of $\theta_{23}$. If inverted, the next generation $0 \nu$ double beta decay experiments will certainly see the events (assuming Majorana neutrinos).

\noindent
(3) The EXO-200 experiment saw $\simeq 10$ $0 \nu$ double beta decay like events over 30 background \cite{Albert:2014awa}. The mass range is consistent with that discussed e.g., in \cite{Battye:2013xqa}.

Do they mean that nature is so kind that she is going to deliver us a ticket for the final World Neutrino Cup in 2024?

\Acknowledgements

At everyday coffee table in the morning, I enjoyed interesting conversations with Francis Halzen, which stimulates my interests and cultivates understanding in the evolving extremely interesting field of high-energy neutrino astrophysics. I thank Renata Zukanovich Funchal for numerous discussions. I express my gratitude to Universidade de S\~ao Paulo for the great opportunity of stay as Pesquisador Visitante Internacional. I am grateful to Conselho Nacional de Ci\^encia e Tecnologia (CNPq) for support for his visit to the Departamento de F\'{\i}sica, Pontif{\'\i}cia Universidade Cat{\'o}lica do Rio de Janeiro, where part of the works in \cite{Machado:2013kya} and \cite{MNQ-Majorana-phase} has been executed. The works described in this article which involve the present author are partially supported by KAKENHI received through Tokyo Metropolitan University, Grant-in-Aid for Scientific Research No. 23540315, Japan Society for the Promotion of Science.

\end{document}